\documentclass[
aps, 
prb, 
10pt,  
amssymb, 
amsmath, 
superscriptaddress, 
tightenlines, 
twocolumn, 
notitlepage, 
] {revtex4-2}

\usepackage{graphicx}
\usepackage{dcolumn}
\usepackage{bm}

\begin{document}


\title{Sublinear transport in Kagome metals:\\ Interplay of Dirac cones and Van Hove singularities}

\author{Nikolai Peshcherenko}%
\affiliation{%
Max Planck Institute for Chemical Physics of Solids, 01187, Dresden, Germany}

\author{Ning Mao}%
\affiliation{Max Planck Institute for Chemical Physics of Solids, 01187, Dresden, Germany}

\author{Claudia Felser}%
\affiliation{Max Planck Institute for Chemical Physics of Solids, 01187, Dresden, Germany}

\author{Yang Zhang}
\email{yangzhang@utk.edu}
\affiliation{Department of Physics and Astronomy, University of Tennessee, Knoxville, TN 37996, USA}
\affiliation{Min H. Kao Department of Electrical Engineering and Computer Science, University of Tennessee, Knoxville, Tennessee 37996, USA}


\begin{abstract}
Kagome metals are known to host Dirac fermions and saddle point Van Hove singularities near Fermi level. With the minimal two-pocket model (Dirac cone + Van Hove singularity), we propose a semiclassical theory to explain the experimentally observed sublinear resistivity in Ni$_3$In and other Kagome metals. We derive the full semiclassical description of kinetic phenomena using Boltzmann equation, and demonstrate that internode electron-electron interaction leads to sublinear in $T$ scaling for both electrical and thermal transport at low temperatures. At higher temperatures above the Dirac node chemical potential, thermal and electric current dissipate through distinct scattering channels, making a ground for Wiedemann-Franz law violation. 
\end{abstract}

\maketitle

\textit{Introduction.} In itinerant metals, electrons/holes are the dominant charge carriers, and the Fermi liquid framework typically applies. Consequently, a clean system is expected to exhibit a resistivity scaling with $T^2$ at intermediate temperatures, reflecting electron-electron interactions and signaling Fermi liquid behavior. The deviations from this well-known result are studied both experimentally and theoretically as the signatures of strongly interacting systems, including Luttinger liquid systems \cite{Luttinger}, heavy fermion metals \cite{heavyfermion}, underdoped cuprates \cite{cuprates}, graphene at charge neutrality point \cite{graphene_wf_law_violation, graphene_th} etc. Non-trivial sublinear temperature behavior was also predicted for materials in the vicinity of quantum phase transition \cite{kineq}. Although these systems have quite different properties, the non-Fermi liquid behaviors are all linked with emerging collective excitations due to strong interactions.

Recently, Kagome lattices have emerged as pivotal platforms for exploring strongly-correlated phenomenon \cite{correlations2,correlations3}, topological magnetism \cite{topological_magnets,magnetism1,magnetism2,magnetism3,magnetism4,magnetism5,magnetism6}, and unconventional superconductivity \cite{correlations1}. Experiment on Kagome metal Ni$_3$In  \cite{the_main_sublinear_reference} has demonstrated a quite unexpected and well-pronounced sublinear in $T$ resistivity at relatively high temperatures $T\gtrsim 100\,\mathrm{K}$, where strong quantum fluctuations or collective excitations are unlikely to appear. 
Similar sublinear behavior, albeit less pronounced, has been observed in three-dimensional Kagome compounds like ScV$_6$Sn$_6$, CsV$_3$Sb$_5$, RbV$_3$Sb$_5$, and KV$_3$Sb$_5$ around room temperature \cite{sublinearexp,sublinear2,sublinear3,sublinear4}. Therefore, the most plausible explanation for this type of behavior should rather be semiclassical. However, existing semiclassical scattering contributions from electron-phonon coupling predict $\rho(T)\propto T^{n}$, $n\ge1$ \cite{abrikosov2017fundamentals}, and impurity scattering is known to give $T$-independent contribution.


In this work, we introduce a two-pocket model consisting of Dirac cone and Van Hove singularity, which allows a sublinear temperature scaling of resistivity reminiscent of non-Fermi liquid like behavior. 
Our model study focuses exclusively on single electron excitations, which are semiclassically described using the Boltzmann equation. The crucial ingredient is the internode electron-electron interaction, which is enhanced due to the divergent density of states from VHS at Fermi level. We demonstrate that, although being momentum-conserving, this scattering process in the presense of momentum-relaxing scattering processes (e.g., impurity scattering) could still provide a leading $T$-sublinear contribution to resistivity. Our theory prediction agrees well with experimental observations \cite{the_main_sublinear_reference,sublinearexp,sublinear1,sublinear2,sublinear3,sublinear4}. In the same framework, the thermal conductivity is also sublinear, though with a different exponent. 

We further demonstrate that the two-pocket model allows for the Wiedemann-Franz law breakdown and non-trivial temperature dependence of Lorentz number. At temperatures higher than Dirac node chemical potential both electrons and holes are thermally activated near the Dirac node. And the intranode scattering in Dirac systems is able to relax electric but not thermal current \cite{graphene_wf_law_violation, schmaliangraphene}. Thus, thermal current is relaxed only with internode scattering.
In this sense, our mechanism is different from previously discussed Wiedemann-Franz law violation in cuprates due to spin-charge separation \cite{spincharge1,spincharge2,spincharge3}. In our setup, the dominant electric and thermal carriers are essentially the same, but they are relaxed by intranode and internode scattering, respectively. 
\begin{figure}
    \centering
    \includegraphics[width=0.5\textwidth]{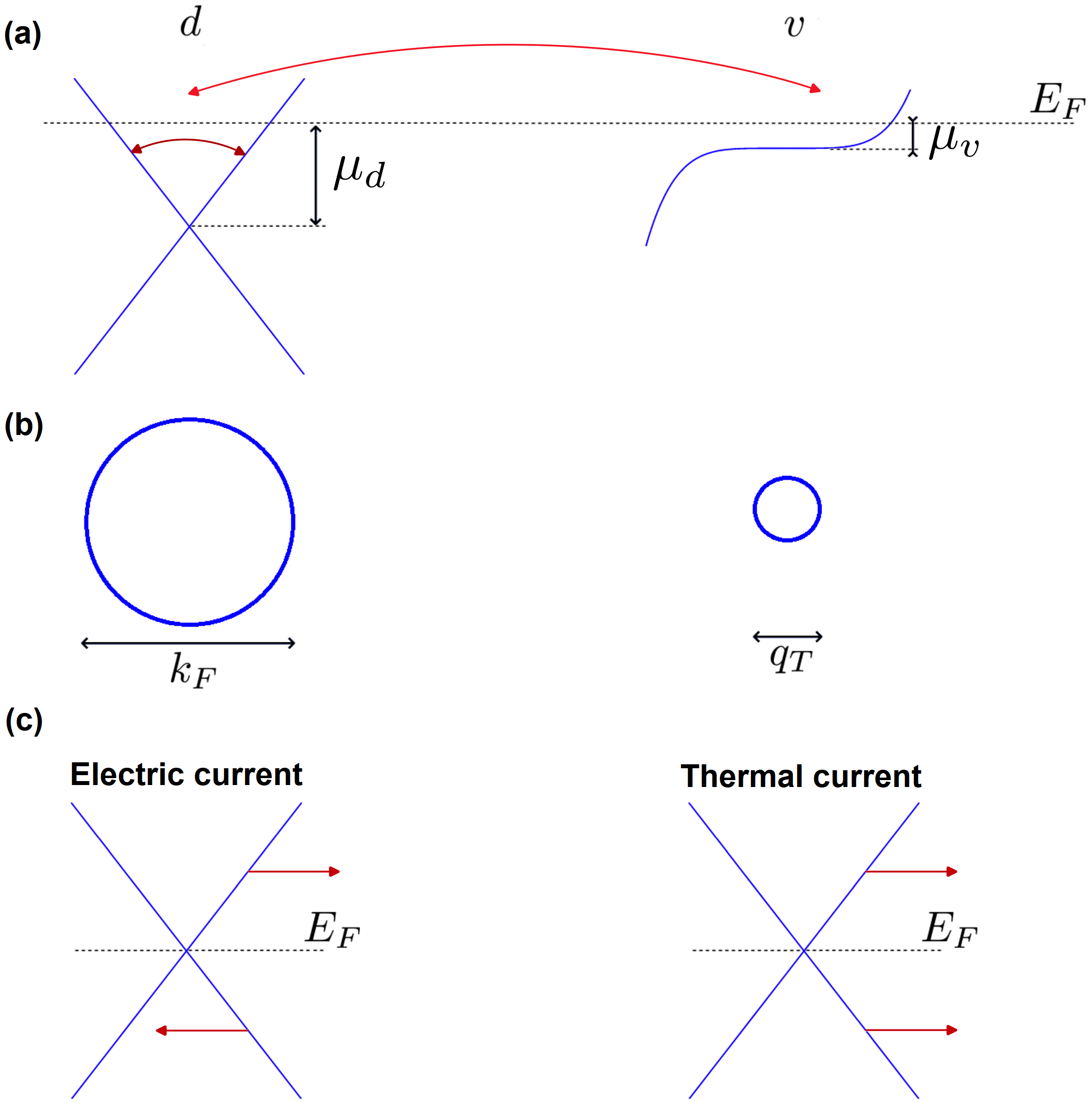}
    \caption{(a) Two-pocket model of a Kagome metal. Node \textit{d} is a Dirac pocket with relatively large Fermi surface. Fast Dirac electrons relax due to both intranode and internode electron-electron scattering. Node \textit{v} has weak dispersion $E_v\sim Ak^{\alpha}$, $\alpha>2$ and possesses small Fermi surface. $\mu_d$, $\mu_v$ are local chemical potential. (b) Fermi pockets for $T\ll\mu_d$. For $T\gg\mu_v$, the size of Fermi surface $q_T$ at node \textit{v} is $T$-dependent: $Aq_T^\alpha\lesssim T$, or, equivalently, $q_T\lesssim\left(T/A\right)^{1/\alpha}$. Due to weak dispersion of VHS electrons, fast Dirac electrons scattering process is almost energy-conserving. This internode process is dominant due to high $\nu_v$ and allows for sublinear $T$ dependence. (c) For $T\gg\mu_d$, intranode scattering is able to relax only charge but not thermal current. Here, an electron-hole pair taken at the same momentum provides opposite in sign contributions to charge transport, but same sign for heat transport. Thus, contribution to heat transport is proportional to total momentum and could not be relaxed by intranode scattering \cite{schmaliangraphene}.}
    \label{fig:2node}
\end{figure}

\textit{Two-pocket model.} The abundant exotic states in Kagome materials are deeply linked to its remarkable band structures. As revealed by density functional theory calculations \cite{kagome2,kagome4,tan2023abundant,Zheng_2024}, ARPES \cite{kagome1,kagome3} and scanning tunneling microscope spectroscopy experiments, the band structure consists of a Dirac point, a saddle point Van Hove singularity (VHS) and a flat band \cite{flatbands1,flatbands2,flatbands3}. We start with explicit introduction of a two-pocket model (see Fig. \ref{fig:2node}), as flat band is generally away from Fermi level. The fast pocket is basically a Dirac pocket with an isotropic Dirac spectrum
\begin{align}
    E_d(\mathbf{k})=v_F|\mathbf{k}|
\end{align}
and the slow pocket has non-trivial momentum dispersion in $k_z=0$ plane:
\begin{align}
    E_v(\mathbf{k})=A\left(|k_x|^{\alpha}-|k_y|^{\alpha}\right),\quad \alpha>2.
\end{align}
We denote two pockets as \textit{d} (Dirac) and \textit{v} (VHS), correspondingly. Density of electronic states from node \textit{v} $\nu_v(E)$ may demonstrate VHS behavior:
\begin{align}
  \nu_v(E)=\sum_\mathbf{k}\delta\left(E-E_v(\mathbf{k})\right)\propto \frac{1}{\alpha A}\left(\frac{A}{|E|}\right)^\varepsilon,\nonumber\\ \varepsilon=1-\frac{2}{\alpha}
  \label{dosvanhove}
\end{align}
for $\alpha>2$. In Kagome metals such as Ni$_3$In and ScV$_6$Sn$_6$, we have the saddle point VHS near $M$ momentum (see discussion in experimental data comparison section and in appendix C). Since \textit{v} electrons are slow (have relatively small Fermi velocities), all transport currents are mainly carried by fast Dirac electrons \textit{d}. 

However, \textit{v} pocket electrons still influence transport properties. This is due to their high density of states which enhances scattering rate of internode electron-electron interactions. For the experimentally relevant case $T\gg\mu_v$ \cite{the_main_sublinear_reference,sublinearexp} ('thermally activated' VHS, for definition of $\mu_v$ see Fig. \ref{fig:2node}a), the effective \textit{v} pocket size $q_T\propto T^{1/\alpha}$ sublinearly depends on $T$ (see description for Fig. \ref{fig:2node}). At the same time, $E_v(\mathbf{q})$ demonstrates relatively weak dispersion. Thus, the scattering of Dirac electrons at \textit{v} node electrons proceeds with negligible energy transfer. This is why \textit{v} electrons could be treated as a momentum-relaxing reservoir with its size $q_T$ dependent on $T$. We show below that the observed  sublinear $\rho(T)$ behavior \cite{the_main_sublinear_reference,sublinearexp,sublinear1,sublinear2,sublinear3,sublinear4} stems from this $T$-dependent \textit{v} pocket size $q_T$.

Now we study in detail the case of relatively large Dirac Fermi surface at $T\ll\mu_d$, so that a standard low-$T$ quasiparticle picture holds. Nevertheless, we also give consideration to $T\gg\mu_d$ limit since it is experimentally feasible and provides nice and non-trivial physics.

\textit{Boltzmann equation description.} Fast current-carrying Dirac electrons distribution function $f_d$ obeys semiclassical Boltzmann equation:
\begin{align}
    \partial_t f_d+\mathbf{v}\cdot\partial_\mathbf{r}f_d+e\mathbf{E}\cdot\partial_\mathbf{p}f_d=I_\mathrm{intra}+I_\mathrm{inter}+\nonumber\\+I_\mathrm{imp}+I_\mathrm{nc},
    \label{masterboltzmann}
\end{align}
where $I_\mathrm{intra}$, $I_\mathrm{inter}$ stand for intra- and internode electron-electron scattering, $I_\mathrm{imp}$ is for impurity scattering and $I_\mathrm{nc}$ describes other types of momentum-relaxing contributions (e.g., scattering at phonons or Umklapp processes for electron-electron interactions).


Below Debye temperature in Kagome metals \cite{sublinear1,sublinear2,sublinear3,sublinear4,sublinearexp,the_main_sublinear_reference} ), e-e interaction plays a major role in transport phenomena. Since in metallic systems electron-electron interaction is screened, we model it in the form of contact interaction:
\begin{align}
    H_\mathrm{int}(\mathbf{r},\mathbf{r}^{\prime})=g\delta\left(\mathbf{r}-\mathbf{r}^{\prime}\right),
    \label{eq:short_range}
\end{align}
with an exception for $T\gg\mu_d$ case (see below). This assumption is justified for typical momentum transfers that are small compared to inversed screening length. Umklapp processes are very rare due to very large momentum transfer required and are therefore omitted.

\begin{figure}
    \centering
    \includegraphics[width=0.45\textwidth]{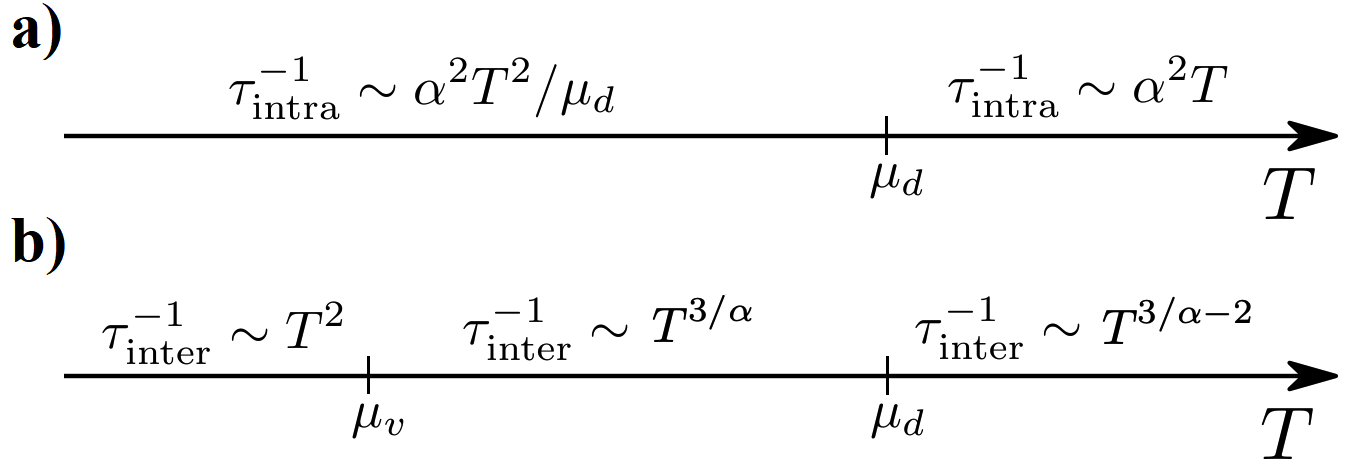}
    \caption{Electron-electron scattering rate as a function of temperature for a) Dirac intranode scattering and b) internode scattering processes. For $T\gg\mu_v$ the size of node \textit{v} Fermi surface is $T$-dependent, which gives rise to sublinear $T$ dependence. For $T\ll\mu_d$ internode scattering is dominant due to high density of states in \textit{v} node. However, for $T\gg\mu_d$ intranode scattering is stronger since intranode interactions are not screened.  Coupling constant $\alpha=e^2/\hbar v_F$ is assumed to be small.}
    
    \label{fig:e-e}
\end{figure}

Electron-electron scattering processes could be basically of two types: intranode (happening between electrons from the same node) or internode (between different nodes), for an outline see Fig. \ref{fig:e-e}. We tackle these two processes separately. 

\textit{Intranode scattering.} Intranode scattering rate of Dirac electrons was previously evaluated in \cite{nonfermi1,nonfermi2,bib:relaxationtime}. We note that intranode processes within \textit{v} node are not relevant to transport properties, since the corresponding electronic states possess low group velocity and thus provide only a small direct contribution to transport currents. Result for electron-electron relaxation time depends on the relation between Dirac fermions chemical potential $\mu_d$ (see Fig. \ref{fig:2node}a) and temperature $T$. For relaxation rate $\tau_\mathrm{intra}^{-1}$ we have:
\begin{align}
    \tau_\mathrm{intra}^{-1}\propto \left(\frac{e^2}{\hbar v_F}\right)^2\begin{cases}
        \frac{T^2}{\mu_d}, & T\ll\mu_d\\
        T, & T\gg\mu_d
    \end{cases},
    \label{eq:fermiliquid}
\end{align}
so that a crossover between Fermi ($\propto T^2$) and non-Fermi liquid behavior happens at $T\sim\mu_d$.

\textit{Internode scattering.} Internode electron-electron scattering, as we show below, is a crucial ingredient for sublinear $T$ behavior of transport coefficients. In appendix A, we explicitly prove that for screened electron-electron interaction \eqref{eq:short_range} $I_\mathrm{inter}$ is given by
\begin{align}
    I_\mathrm{inter}=-\frac{\delta f_d(\mathbf{p})}{\tau_\mathrm{inter}},\nonumber\\
    \frac{1}{\tau_\mathrm{inter}}=2\pi g^2\int\frac{d^2\mathbf{q}}{(2\pi)^2}F(\mathbf{q},T)\delta\left(E_d(\mathbf{p})-E_d(\mathbf{p}+\mathbf{q})\right)\cdot\nonumber\\\left(1-\cos\theta_{\mathbf{p+q},\mathbf{p}}\right),\nonumber\\F(\mathbf{q},T)=\sum_\mathbf{p_v}f_v(\mathbf{p_v})\left(1-f_v(\mathbf{p_v-q})\right)
\end{align}
where $\mathbf{p_v}$ and $f_v$ are VHS electrons momentum and distribution function correspondingly. Initial momentum of Dirac electron is $\mathbf{p}$ and $\mathbf{q}$ stands for momentum transfer. Detailed evaluation of $F(\mathbf{q},T)$ and $I_\mathrm{inter}$ performed in appendices \ref{app:Icollder}, \ref{ap:F} for $T\ll\mu_d$ gives
\begin{align}
    \frac{1}{\tau_\mathrm{e-e}}\sim\frac{g^2n_v}{\mu_d^2/v_F}\left(\frac{T}{A}\right)^{3/\alpha},
    \label{eq:sublineartau}
\end{align}
where $n_v$ is VHS electrons concentration. Therefore, high-order VHS-mediated internode scattering leads to sublinear scaling with $T$. 

\textit{Momentum-relaxing processes and sublinear $T$ behavior.} However, momentum-conserving electron-electron interactions alone would lead to non-decaying current, thus a minimal transport estimation of Kagome metals should include momentum-relaxing processes. As experimental estimates for ScV$_6$Sn$_6$ \cite{sublinearexp} show, impurity scattering exhibits the shortest scattering time among momentum-relaxing processes. Electron-phonon interactions prove to be negligible \cite{sublinearexp}, since sublinear $T$ behavior is observed below Debye temperature. Thus, the minimal relaxation term should include internode electron-electron scattering and impurity relaxation. 

\begin{figure}
    \centering
    \includegraphics[width=0.47\textwidth]{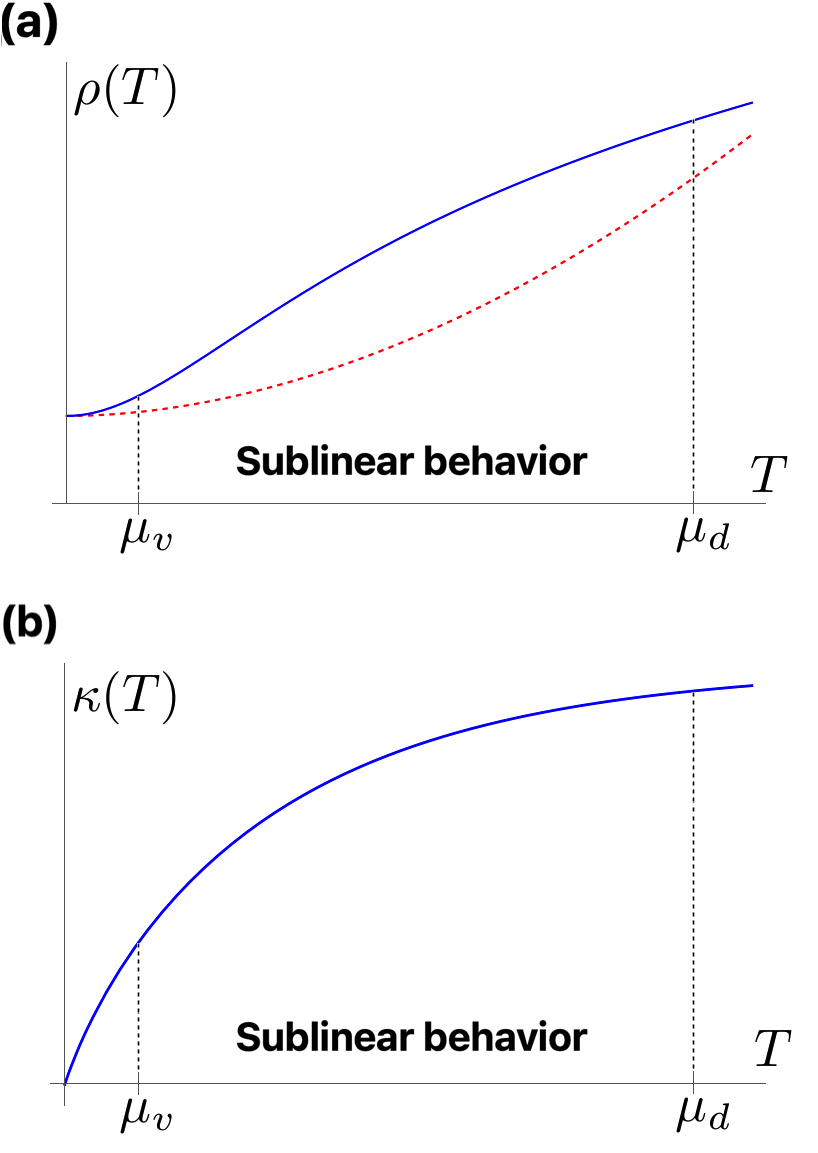}
    \caption{(a) $T$-dependent resistivity $\rho(T)$ for the two-pocket model of a Kagome metal (solid line) and a standard Fermi liquid behavior (dashed line). Kagome metals' resistivity demonstrates sublinear temperature behavior $\rho(T)\propto T^{3/\alpha}$ in a range $\mu_v\ll T\ll\mu_d$ due to dominant internode electron-electron scattering rate given by Eq. \eqref{eq:sublineartau}. For large $T\gg\mu_d$, intranode interaction becomes the strongest due to lack of screening and thus $\rho(T)\propto T$. (b) $T$-dependent thermal conductivity $\kappa(T)$. Since for $T\ll\mu_d$ Wiedemann-Franz law holds, $\kappa(T)$ also has a sublinear behavior as $\kappa(T)\propto T^{1-3/\alpha}$ in the range $\mu_v\ll T\ll\mu_d$. For $T\gg\mu_d$, however, Wiedemann-Franz law breaks down due to separation of electric and thermal current relaxation channels (see discussion below).}
    \label{fig:res}
\end{figure}

Now, following a well-known textbook procedure \cite{Landau_kinetics} for finding kinetic coefficients, one can see that Boltzmann equation \eqref{masterboltzmann} transforms into
\begin{align}
    \mathbf{v}\cdot\left[e\mathbf{E}-\frac{\varepsilon-\mu_d}{T}\nabla T\right]\partial_\varepsilon f_\mathrm{eq}=-\frac{\delta f_d}{\tau_\mathrm{imp}}-\frac{\delta f_d}{\tau_\mathrm{inter}}
\end{align}
so that
\begin{align}
    \delta f_d=e\tau_\mathrm{eff}\left(-\partial_\varepsilon f_\mathrm{eq}\right)\mathbf{v}\cdot\left[e\mathbf{E}-\frac{\varepsilon-\mu_d}{T}\nabla T\right],\nonumber\\ \tau_\mathrm{eff}=\left(\tau_\mathrm{imp}^{-1}+\tau_\mathrm{inter}^{-1}\right)^{-1}
\end{align}
and for charge and heat currents $\mathbf{j}$, $\mathbf{q}$ of Dirac electrons we arrive at
\begin{align}
    \mathbf{j}=e^2\sum_\mathbf{p}\mathbf{v}_d\left(\mathbf{v}_d\mathbf{E}\right)\tau_\mathrm{eff}\left(-\partial_\varepsilon f_\mathrm{eq}\right)
    \label{jd},
\end{align}
\begin{align}
    \mathbf{q}=-\sum_\mathbf{p}\frac{(\varepsilon_d-\mu_d)^2}{T}\mathbf{v}_d\left(\mathbf{v}_d\nabla T\right)\tau_\mathrm{eff}\left(-\partial_\varepsilon f_\mathrm{eq}\right),
    \label{qd}
\end{align}
which for $T\ll\mu_d$ gives
\begin{align}
    \sigma=e^2\nu_d(\mu_d)D,\quad \kappa=\frac{\pi^2}{3}\nu_d(\mu_d)DT,\nonumber\\ D=\frac{1}{2}v_F^2\tau_\mathrm{eff}(\mu_d,T).
    \label{results}
\end{align}
We plot the resistivity and thermal conductivity at Fig. \ref{fig:res}. Despite demonstrating non Fermi liquid like behavior of sublinear $T$-dependence, Eq. \eqref{results} suggests that Wiedemann-Franz law still holds. This is because both charge and heat currents decay due to the same process, namely, internode electron-electron scattering. However, as we discuss later, for high temperatures $T\gg\mu_d$ it is not so and Wiedemann-Franz law breaks down. 

\textit{Comparison with experimental data.} As mentioned in introduction, the well-pronounced sublinear in temperature resistivity was observed \cite{the_main_sublinear_reference} in Ni$_3$In compound for $T\gtrsim 100$ K. In the subsequent sections, we illustrate how our two-pocket semiclassical model offers a plausible explanation for this phenomenon.

\begin{figure}
    \centering
    \includegraphics[width=0.5\textwidth]{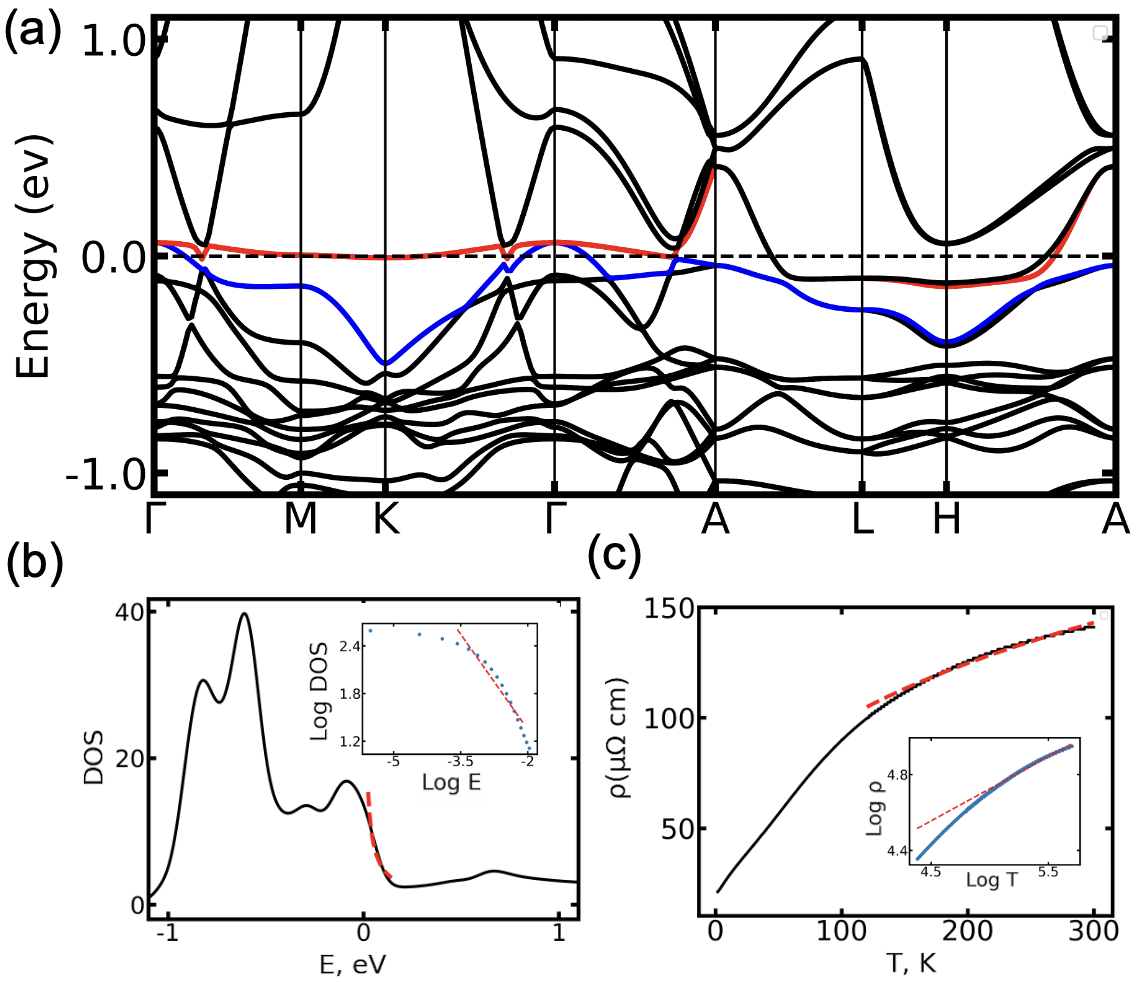}
    \caption{(a) DFT electronic structure of Ni$_3$In. Fermi level is shown by dashed line. (b) DFT total density of states shows a peak near Fermi level. Dashed red line stands for the fitting as $(E-E_\mathrm{VHS})^{-\beta}$, $\beta=0.8$ (please see inset for log log plot), $E_\mathrm{VHS}=0$. (c) Ni$_3$In experimental resistivity $\rho(T)$ \cite{the_main_sublinear_reference} versus temperature is shown. Sublinear fitting of Ni$_3$In resistivity $\rho(T)\propto T^{\gamma}$ gives $\gamma_\mathrm{exp}=0.34$ (red dashed line and inset log-log plot) in accord with our theoretical prediction.}
    \label{fig:NiIn}
\end{figure}

Band structure calculation for Ni$_3$In (see Fig. \ref{fig:NiIn}a) identifies the fast Dirac pocket near $\Gamma$ point. Further, the DFT density of states (see Fig. \ref{fig:NiIn}b) demonstrates a peak around Fermi energy. Density of states power law fitting DOS$(\varepsilon)\propto(E-E_\mathrm{VHS})^{-\beta}$ predicts $\beta=0.8$, a power-law divergent higher order Van Hove singularity \cite{supermetal}. According to Eq. \eqref{dosvanhove}, \eqref{eq:sublineartau}, this predicts resistivity exponent $\gamma_\mathrm{th}=\frac{3}{2}(1-\beta)=0.3$ ($\rho(T)\propto T^\gamma$). This agrees well with experimental data fitting result $\gamma_\mathrm{exp}=0.34$. We note the suggested mechanism does not describe resistivity curve Fig.\ref{fig:NiIn}c below $\sim 100$ K. As discussed in Ref\cite{the_main_sublinear_reference}, below $100$ K, the resistivity is dominated by a strongly non-Fermi liquid correlations. However, for higher temperatures this correlated state is claimed to be destroyed, falling into the semiclassical regime.
We further perform DFT calculations for other Kagome compounds such as ScV$_6$Sn$_6$, CsV$_3$Sb$_5$, RbV$_3$Sb$_5$, KV$_3$Sb$_5$. In appendix C, we prove with the help of kp model fitting that experimentally observed \cite{sublinearexp} universal sublinear resistivity behavior agrees well with our theory prediction. 

\textit{Wiedemann-Franz law break down.} Another interesting aspect of the two-pocket model is Wiedemann-Franz law violation at high temperatures case ($T \ge \mu_d$). 
The electron-hole symmetry is approximately restored within this limit for Dirac node, which essentially implies the absence of screening for intranode interactions (screening length \cite{screening} $l^2_\mathrm{scr}\propto1/\nu(\mu_d)\to\infty$).
This in turn makes the quantitative theoretical description of transport phenomena cumbersome. Hence, the aim of this section is not to give accurate derivations but rather reasonable estimates for the behavior of Lorentz number $L(T)$ with temperature. For $L(T)$, one could write
\begin{align}
    \frac{L(T)}{L_\mathrm{FL}}\sim\frac{\tau_\kappa}{\tau_\mathrm{intra}},
    \label{eq:Lorentz_estimate}
\end{align}
where $\tau_\kappa=\mathrm{min}\{\tau_\mathrm{inter},\,\tau_\mathrm{imp}\}$ describes thermal current relaxation time. According to Ref.\cite{schmaliangraphene}, intranode electron-electron scattering is unable to relax thermal current (see also Fig. \ref{fig:2node} c). It is therefore done by internode and impurity scattering. Charge current relaxation time is still given by the shortest time scale (which is now $\tau_\mathrm{intra}$ due to the absence of screening under approximate electron-hole symmetry). The separation of thermal and charge currents relaxation channels is the ultimate reason for Wiedemann-Franz law break down in this limit. For the schematic plot of $L(T)$ behavior please see appendix E.

 
\textit{Conclusions.} In this work, we have proposed a semiclassical theory for non-Fermi liquid-like sublinear charge and heat transport coefficients, based on a two-pocket model (Dirac node + Van Hove singularity) of Kagome metals. When Fermi level is sufficiently close to VHS, the dominant scattering source of current-carrying Dirac electrons is internode electron-electron scattering. We have demonstrated that electronic states near high order Van Hove singularity effectively form a momentum relaxing reservoir for fast Dirac electrons, allowing only for a vanishing energy exchange. The reservoir size is though $T$-dependent: $q_\mathrm{VHS}\lesssim\left(T/A\right)^{1/\alpha}$. Combined with relatively large Dirac pocket ($q_D\gg q_\mathrm{VHS}$) or generic fast electron pocket, this allows for a sublinear $T$-dependence of transport coefficients, $\rho(T)\propto T^{3/\alpha}$, $\kappa\propto T^{1-3/\alpha}$. Our finding for resistivity scaling directly explains the recent transport experiments in Kagome metal \cite{the_main_sublinear_reference,sublinearexp}. Nevertheless, despite the non-Fermi liquid like behavior at relatively low temperatures, Wiedemann-Franz law still holds since both charge and heat currents relax due to the same internode scattering process. While for relatively large temperatures, the thermal and electric current relaxation channels split due to the approximate electron-hole symmetry of Dirac node \cite{schmaliangraphene}, leading to Wiedemann-Franz law violation.

\textbf{Acknowledgement.}
We are grateful to Hiroki Isobe and Liang Fu for helpful discussions, Shirin Mozaffari and David Mandrus for related experimental collaborations. C.F. is supported by the Catalyst Fund of Canadian Institute for Advanced Research. Y. Z. is supported by the start-up fund at University of Tennessee Knoxville. 

\nocite{*}

\bibliography{sample}

\appendix

\section{Internode scattering time evaluation\label{app:Icollder}}
\begin{widetext}
    In this appendix we provide a semiclassical derivation of electron-electron scattering rate within the Boltzmann equation framework. Kinetic equations for distribution functions of fast Dirac electrons $f_d$ and slow VHS electrons $f_v$ are given by
\begin{align}
    \partial_t f_d+\mathbf{v}\cdot\partial_\mathbf{r}f_d+e\mathbf{E}\cdot\partial_\mathbf{p}f_d=I^d_{\mathrm{imp}}\left[f_d\right]+I^d_{\mathrm{e}}\left[f_d,\,f_v\right]\label{boltzmann1}\\
        \partial_t f_v+\mathbf{v}\cdot\partial_\mathbf{r}f_v+e\mathbf{E}\cdot\partial_\mathbf{p}f_v=I^v_{\mathrm{imp}}\left[f_v\right]+I^v_{\mathrm{e}}\left[f_d,\,f_v\right]
    \label{boltzmann2}
\end{align}
Here $I_\mathrm{imp}$ describes impurity scattering, $I_\mathrm{e}$ - electron-electron scattering.
The crucial ingredient of our theory lies in the electron-electron scattering part of collision integral $I_\mathrm{e}$:
\begin{align}
    I_\mathrm{e}^d\left[f-d,\,f_v\right]=I_\mathrm{e}^d\left[f_d,\,f_v\right]_\mathrm{in}-I_\mathrm{e}^d\left[f_d,\,f_v\right]_\mathrm{out},\nonumber\\ I_\mathrm{e}^d\left[f_d,\,f_v\right]_\mathrm{in}=\sum_{\mathbf{p}^{\prime}}W_{\mathbf{p}^{\prime}\to \mathbf{p}}\, f_d(\mathbf{p}^{\prime})\left(1-f_d(\mathbf{p})\right),\quad I_\mathrm{e}^d\left[f_d,\,f_v\right]_\mathrm{out}=\sum_{\mathbf{p}^{\prime}}W_{\mathbf{p}\to\mathbf{p}^{\prime}}\cdot f_d(\mathbf{p})\left(1-f_d(\mathbf{p}^{\prime})\right).
    \label{intcoll1}
\end{align}
In turn, transition probabilities $W_{\mathbf{p}\to\mathbf{p}^{\prime}}$, $W_{\mathbf{p}^{\prime}\to\mathbf{p}}$ are given by
    \begin{align}
    W_{\mathbf{p}^{\prime}\to \mathbf{p}}=2\pi g^2\sum_{\mathbf{p_1},\mathbf{p_2}}f_v(\mathbf{p_1})\left(1-f_v(\mathbf{p_2})\right)\delta\left(E_d(\mathbf{p})+E_v(\mathbf{p_2})-E_d(\mathbf{p}^{\prime})-E_v(\mathbf{p_1})\right)\delta\left(\mathbf{p}+\mathbf{p}_2-\mathbf{p}^{\prime}-\mathbf{p}_1\right),\nonumber\\
    W_{\mathbf{p}\to \mathbf{p}^{\prime}}=2\pi g^2\sum_{\mathbf{p_1},\mathbf{p_2}}f_v(\mathbf{p_2})\left(1-f_v(\mathbf{p_1})\right)\delta\left(E_d(\mathbf{p}^{\prime})+E_v(\mathbf{p_1})-E_d(\mathbf{p})-E_v(\mathbf{p_2})\right)\delta\left(\mathbf{p}^{\prime}+\mathbf{p}_1-\mathbf{p}-\mathbf{p}_2\right).
    \label{intcoll2}
\end{align}

In \eqref{intcoll1}, \eqref{intcoll2} $\mathbf{p},\mathbf{p}^{\prime}$ are initial and final momenta of a Dirac electron, $\mathbf{p_1},\mathbf{p_2}$ are the same but for VHS electron. The equation \eqref{intcoll2} itself is very general and can describe, for instance, the standard $T^2$ metal behavior either for intra- or interpocket scattering processes (including Umklapp processes). However, in Kagome metal case $E_v(\mathbf{p})$ demonstrates a very weak dispersion, meaning that one can adopt $E_v(\mathbf{p}_{1})\approx E_v(\mathbf{p}_{2})$. 

In order to make it easier to compare with calculations of previous sections change momentum variables from $\mathbf{p}_1$, $\mathbf{p}_2$, $\mathbf{p^{\prime}}$ to the following ones:
\begin{align}
    \mathbf{p^{\prime}}=\mathbf{p}+\mathbf{q},\quad \mathbf{p}_2=\mathbf{p}_v-\mathbf{q},\quad \mathbf{p}_v=\mathbf{p}_1.
\end{align}
In these notations momentum conservation law is fullfiled automatically. For transition probabilities from \eqref{intcoll2} we then arrive at
\begin{align}
    W_{\mathbf{p}^{\prime}\to \mathbf{p}}\approx 2\pi g^2 \sum_{\mathbf{p}_v}f_v(\mathbf{p}_v)\left(1-f_v(\mathbf{p}_v+\mathbf{q})\right)\delta\left(E_d(\mathbf{p})-E_d(\mathbf{p}+\mathbf{q})\right),\nonumber\\W_{\mathbf{p}\to \mathbf{p}^{\prime}}\approx 2\pi g^2 \sum_{\mathbf{p}_v}f_v(\mathbf{p}_v)\left(1-f_v(\mathbf{p}_v-\mathbf{q})\right)\delta\left(E_d(\mathbf{p})-E_d(\mathbf{p}+\mathbf{q})\right)
\end{align}
Since for equilibrium distribution function $f_v(\mathbf{k})=f_v(\mathbf{-k})$, we find in accord with detailed balance principle
\begin{align}
    W_{\mathbf{p}^{\prime}\to \mathbf{p}}= W_{\mathbf{p}\to \mathbf{p}^{\prime}}= 2\pi g^2 F(\mathbf{q},T) \delta\left(E_d(\mathbf{p})-E_d(\mathbf{p}+\mathbf{q})\right),\nonumber\\ F(\mathbf{q},T)\equiv\sum_{\mathbf{p}_v}f_v(\mathbf{p}_v)\left(1-f_v(\mathbf{p}_v-\mathbf{q})\right)=\int\frac{d^2\mathbf{p}_v}{(2\pi)^2}\frac{1}{\left(e^{\beta(E_v(\mathbf{p}_v)-\mu_v)}+1\right)\left(e^{-\beta(E_v(\mathbf{p}_v-\mathbf{q})-\mu_v)}+1\right)}
\end{align}
so that
\begin{align}
    I_e^d\left [f_d\right]=\sum_{\mathbf{q}}W_{\mathbf{p}+\mathbf{q}\to \mathbf{p}}\cdot f_d(\mathbf{p}+\mathbf{q})\left(1-f_d(\mathbf{p})\right)-\sum_{\mathbf{q}}W_{\mathbf{p}\to\mathbf{p}+\mathbf{q}}\cdot f_d(\mathbf{p})\left(1-f_d(\mathbf{p}+\mathbf{q})\right)=\nonumber\\=2\pi g^2\sum_{\mathbf{q}}F(\mathbf{q},T) \delta\left(E_d(\mathbf{p})-E_d(\mathbf{p}+\mathbf{q})\right)\left(f_d(\mathbf{p+q})-f_d(\mathbf{p})\right).
\end{align}
Substituting $f_d(\mathbf{p})=f_\mathrm{eq}(|\mathbf{p}|)+\delta f(\mathbf{p})$ into \eqref{intcoll1} and evaluating the integral in polar coordinates, we arrive at:
    \begin{align}
    I_e^d\left [f_d\right]=g^2\int_0^{2\pi}\frac{d\varphi}{2\pi}\int qdq\,\delta\left[v_F\left(p_F-\sqrt{p_F^2+q^2+2p_Fq\cos\varphi}\right) \right]F(\mathbf{q},T)\left(\delta f(\mathbf{p}+\mathbf{q})-\delta f(\mathbf{p})\right)=\nonumber\\=\frac{g^2}{2\pi v_F}\int dq\frac{\theta(2p_F-q)}{\sqrt{1-q^2/4p_F^2}}F(q,\varphi_q,T)\left(\delta f(\mathbf{p}+\mathbf{q})-\delta f(\mathbf{p})\right)|_{\varphi=\varphi_q}.
\end{align}
As we demonstrate in appendix \ref{ap:F}, $F(\mathbf{q})$ effectively cuts the integral over $q$ at $q_T=(T/A)^{1/\alpha}$. Hence, assuming large thermally activated Fermi surface $q_T\ll k_F$ (which was proven to be experimentally relevant in \cite{sublinearexp}) and a standard Ansatz for $\delta f(\mathbf{p})=\frac{\mathbf{E}\mathbf{p}}{|\mathbf{p}|}\chi(|\mathbf{p}|)$, we arrive at
\begin{align}
    I_e^d\left [f_d\right]=-\frac{\chi}{\tau_\mathrm{e-e}},\quad \frac{1}{\tau_\mathrm{e-e}}\sim\frac{g^2 n_v}{\mu_d^2/v_F}\left(\frac{T}{A}\right)^{3/\alpha}.
\end{align}
Here $n_v$ is the total VHS electrons concentration.
\end{widetext}

\section{Estimate of $F(\mathbf{q},T)$\label{ap:F}}
\begin{figure}
    \centering
    \includegraphics[width=0.45\textwidth]{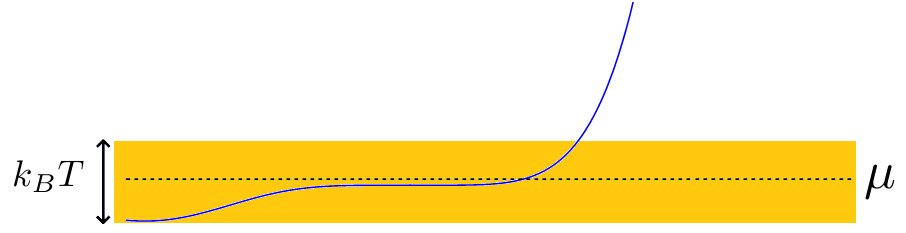}
    \caption{A sketch of Van Hove singularity in Ni$_3$In. Yellow: thermally activated electrons. For high temperatures ($T\gtrsim 100\,\mathrm{K}$) number of activated electrons does not depend on $T$.}
    \label{fig:VHS_saddle}
\end{figure}

In this appendix we give an estimate for temperature dependence of $F(\mathbf{q})$ function given by
\begin{align}
    F(\mathbf{q},T)\equiv\nonumber\\\int\frac{d^2\mathbf{p}_v}{(2\pi)^2}\frac{1}{\left(e^{\beta(E_v(\mathbf{p}_v)-\mu)}+1\right)\left(e^{-\beta(E_v(\mathbf{p}_v-\mathbf{q})-\mu)}+1\right)}.
    \label{eq:F_exact}
\end{align}
Let us note that outside of the region $q\le q_T$ $F(\mathbf{q})$ should decay exponentially. Having in mind Ni$_3$In case (see Fig. 4a of the main text), it is easy to see that for $q\le q_T$ and $T\gg\mu_v$ VHS electrons should obey Boltzmann statistics (see also Fig.\ref{fig:VHS_saddle}). Hence,
\begin{align}
    F(\mathbf{q},T)=\sum_{\mathbf{p}_v}f_v(\mathbf{p}_v)\left(1-f_v(\mathbf{p}_v-\mathbf{q})\right)\approx\nonumber\\\approx\theta\left(|\mathbf{q}|-q_T\right)\sum_{\mathbf{p}_v}f_v(\mathbf{p}_v)=n_v\theta\left(|\mathbf{q}|-q_T\right)
    \label{eq:F_estimate}
\end{align}
Therefore, one can say that for Ni$_3$In material case $F(\mathbf{q},T)$ is actually $T$-independent and gives Van Hove electrons concentration.

Although having an ideal saddle point-type of Van Hove singularity (i.e., uncut from below, see band structures of appendix C) would most probably result in having a $T$-dependent $n_v$, the details of this dependence would be determined by electron bands behavior for energies well below the Fermi surface. Given that the bands demonstrate sharp decay, one could also expect weak $T$-dependence of $n_v$ for other Kagome materials.

\section{Real material data comparison: ScV$_6$Sn$_6$, CsV$_3$Sb$_5$, RbV$_3$Sb$_5$ and KV$_3$Sb$_5$\label{kp_model}}

\begin{figure}[t]
\includegraphics[width=\columnwidth]{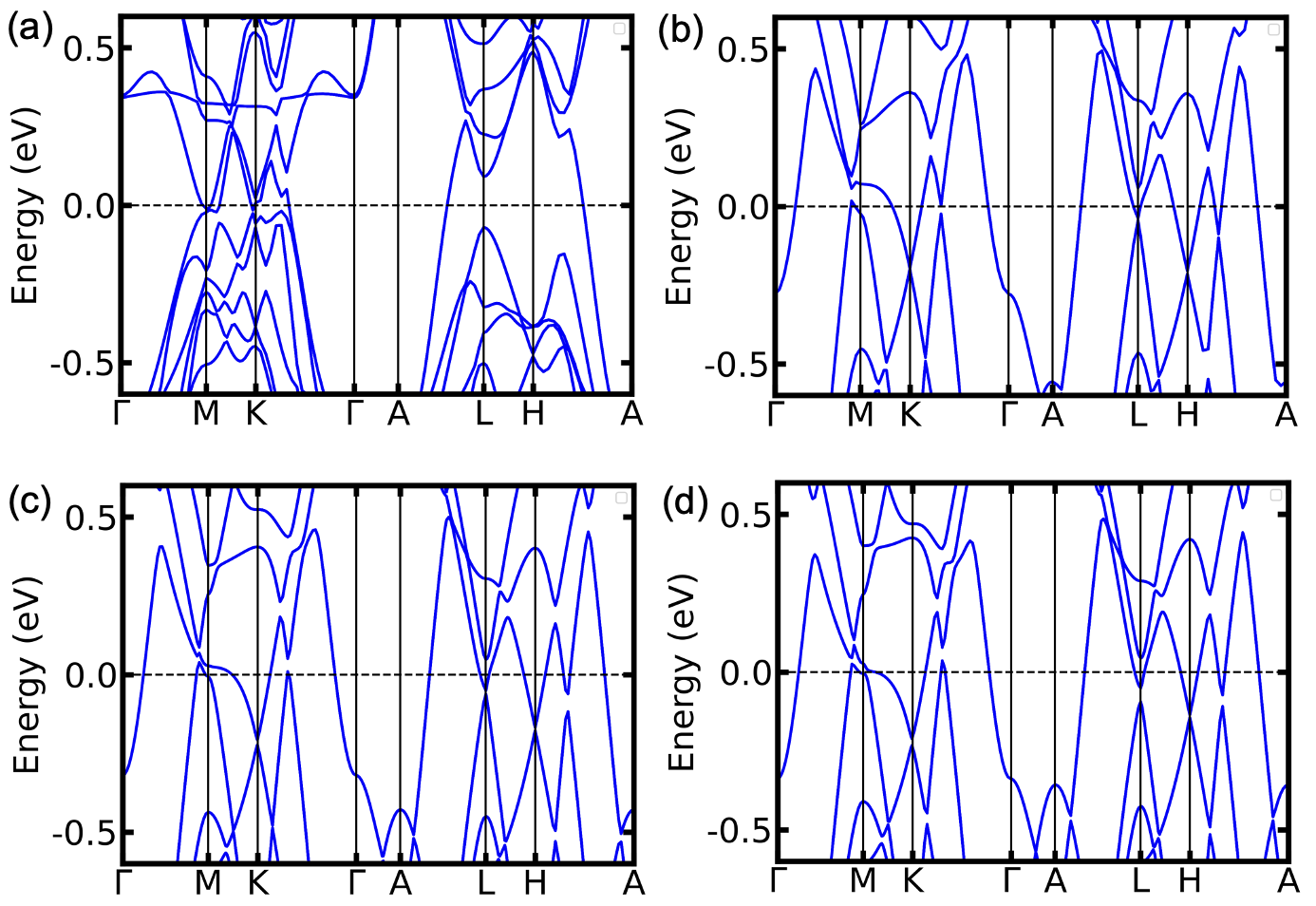}
\caption{ The band structure of (a) ScV$_6$Sn$_6$, (b) CsV$_3$Sb$_5$, (c) RbV$_3$Sb$_5$ and (d) KV$_3$Sb$_5$. All materials listed demonstrate having a fast Dirac pocket near K point and a saddle point type of Van Hove singularity around M point.
}\label{fig_bands}
\end{figure}

In addition to the Ni$_3$In experiments mentioned in the main text, in this appendix we give consideration to other Kagome metals that demonstrated \cite{sublinear1,sublinear2,sublinear3,sublinear4,sublinearexp} similar sublinear scaling of resistivity with temperature $\rho(T)\propto T^\gamma$, $\gamma=0.62$. From VASP-calculated band structure for ScV$_6$Sn$_6$, CsV$_3$Sb$_5$, RbV$_3$Sb$_5$, KV$_3$Sb$_5$ (see Fig. \ref{fig_bands}) one can see that all of these materials have a fast Dirac pocket near K point and a saddle point type of Van Hove singularity around M point, thus making a ground for a simple two-pocket approach developed in the main text.  

\vspace{0.5cm}
\begin{figure}[t]
\includegraphics[width=\columnwidth]{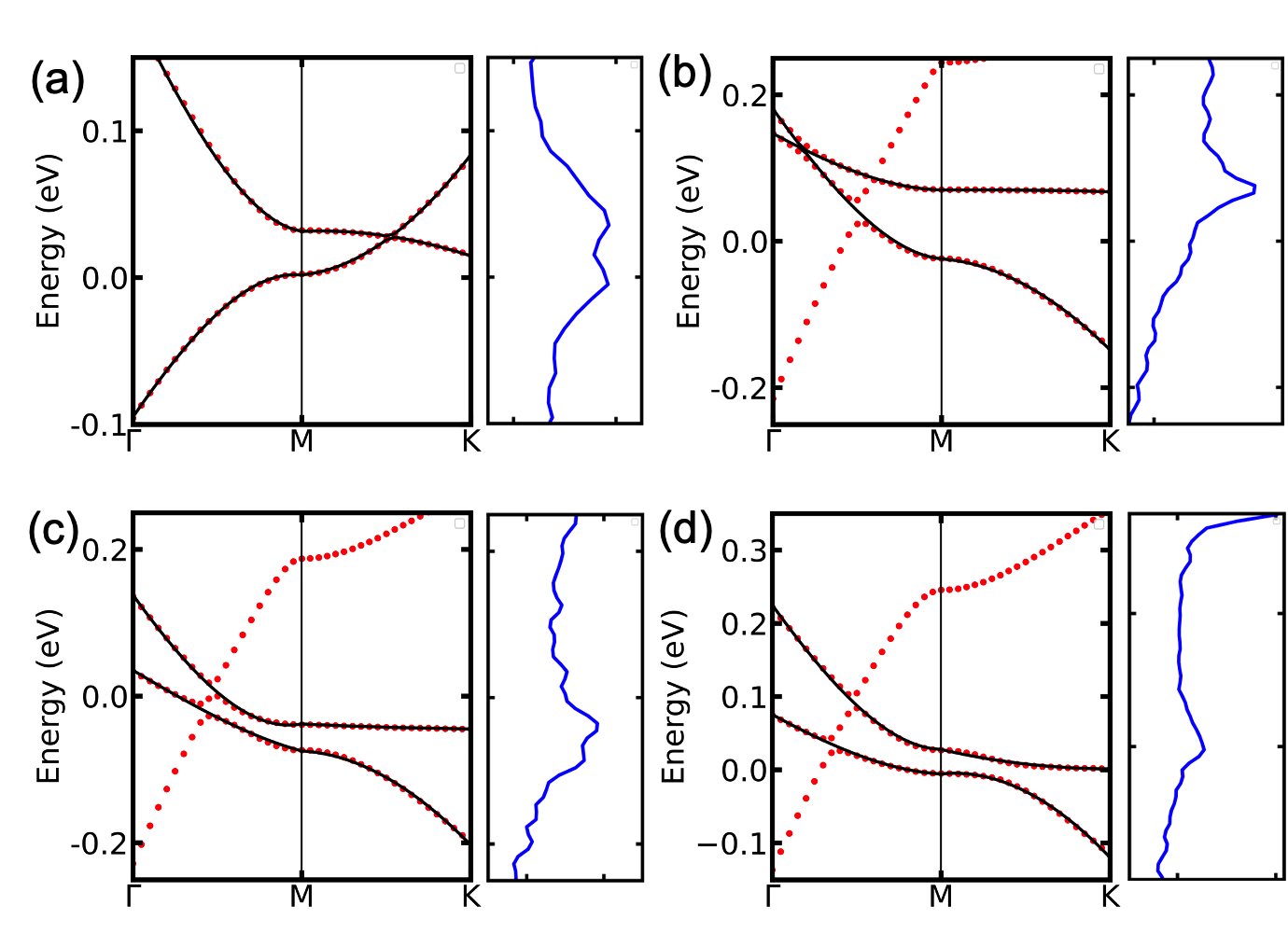}
\caption{
The fitted band Structure of (a) ScV$_6$Sn$_6$, (b) CsV$_3$Sb$_5$, (c) RbV$_3$Sb$_5$, and (d) KV$_3$Sb$_5$. The black lines represent the $k\cdot p$ model's prediction, while the red dots indicate the VASP-calculated band structures in the vicinity of the M points. Notably, there is a switch in the representations around the Dirac point. Calculated density of states is shown by blue line. One can see that in all cases it has a peak near Fermi level.
}\label{fig:kpfit}
\end{figure}

Our prediction for resistivity scaling exponent (see Eq. (8) of the main text) heavily depends on the bands behavior near M point. Thus, in order to compare it with experimental results \cite{sublinearexp} we performed $k\cdot p$ model fitting for the two DFT bands closest to the Fermi level near M point (see Fig. \ref{fig:kpfit}).

We fit the bands near the Fermi level using two polynomial functions, each with a maximum order of five. These polynomials are represented as:
\begin{equation}
H(\mathbf{k}) = 
\begin{pmatrix}
P_1(\mathbf{k}) & 0 \\
0 & P_2(\mathbf{k}) 
\end{pmatrix},
\end{equation}
where $P_1(\mathbf{k})$, $P_2(\mathbf{k})$ are given by
\begin{equation}
    \begin{array}{l} 
          P_1(\mathbf{k}) = a_1 k_x^5  + a_2 k_x^3 + a_3  k_x k_y^2 + a_4 k_y k_x^2  + a_5 \\
          P_2(\mathbf{k}) = b_1 k_x^5  + b_2 k_x^3 + b_3  k_x^2 k_y + b_4 k_x k_y^2 + b_5.
    \end{array}
    \label{eq:fit}
\end{equation}
Please note that although these compounds have $\mathrm{P6/mmm}$ symmetry group, for a two-band modelling not all of the group symmetries could be respected. Thus, we keep inversion symmetry that allows only for odd powers of $k$. By fitting the band structures from VASP calculations with \eqref{eq:fit} we can obtain the 10 expansion coefficients (please see Table \ref{table:fit}). 
\begin{widetext}
\begin{center}
\begin{table}[h!]
\centering    
\begin{tabular}{ |c|c|c|c|c|c|c|c|c|c|c| } 
\hline
Material & $a_1$ & $a_2$ & $a_3$ & $a_4$ & $a_5$ & $b_1$ & $b_2$ & $b_3$ & $b_4$ & $b_5$ \\
\hline
ScV$_6$Sn$_6$ & 0.00319 & -0.0533 & -0.0789 & -0.000259 & 0.710 & -0.00248 & 0.0402 & 0.00176 & 0.168 & -0.484\\ 
\hline
CsV$_3$Sb$_5$ & 0.00170 & -0.0283 & -0.0274 & -0.0000193 & 0.425 & 0.00434 & -0.0726 & -0.00580 & -0.265 & 0.897\\
\hline
RbV$_3$Sb$_5$ & 0.00139 & -0.0261 & -0.247 & -0.00788 & 0.311 & 0.00493 & -0.0787 & 0.000792 & -0.0497 & 0.893 \\ 
\hline
KV$_3$Sb$_5$ & 0.00167 & -0.0281 & -0.281 & 0.00472 & 0.354 & 0.00446 & -0.0736 & -0.0136 & -0.0135 & 0.947\\
\hline
\end{tabular}
\caption{Result of a corresponding material DFT bands fitting with Eq. \eqref{eq:fit}.}
\label{table:fit}
\end{table}
\end{center}
\end{widetext}

In 3D material case power-law type of VHS (see calculated density of states at Fig. \ref{fig:kpfit}) could be provided only by 5th order term. Thus, our DFT bands fitting, according to Eq. (8) of the main text, predicts $0.6$ resistivity exponent which stands in full agreement with experimental value $n=0.62$ from \cite{sublinearexp}.

\section{Anomalous Hall effect\label{stoner}}


Yet another physical aspect a high order Van Hove singularity could bring to transport features is anomalous Hall response. This phenomenon was recently observed experimentally in various Kagome materials \cite{AHE_kagome1,AHE_kagome2,AHE_kagome3,AHE_kagome4}, where it was attributed either to intrinsic (Berry curvature-related) or extrinsic (skew scattering) contributions. However, first-principle calculations conducted for KV$_3$Sb$_5$ \cite{first_principal_AHE} predicts that an intrinsic contribution is much smaller than the one observed in experiment \cite{AHE_kagome2}. And it has been shown \cite{KVSb_magnetic_test} that KV$_3$Sb$_5$ does not support any net magnetic moment so that there is no ground for skew scattering. Hence, other mechanisms should be considered.

We suggest that the role of magnetic impurities could be played by Van Hove singularity electrons. Namely, we imply that Stoner instability for VHS pocket electrons may occur due to diverging density of states. Thus, scattering of fast Dirac electrons at VHS states would allow for a skew-scattering contribution off the electrons magnetic moment.

One can estimate the effective electron exchange constant $U$ to be of the order of $U\sim\nu^{-1}_v$, hence $\tau_\mathrm{skew}$ is given by
\begin{align}
    \frac{1}{\tau_\mathrm{skew}}\sim\frac{1}{\tau_\mathrm{tr}}U\nu_d(\mu)\sim\frac{1}{\tau_\mathrm{tr}}\frac{\nu_d(\mu)}{\nu_v(\mu)},
\end{align}
so that
\begin{align}
    \frac{\rho_{H}}{\rho_0}\sim\frac{\tau_\mathrm{tr}}{\tau_\mathrm{skew}}
\end{align}
where $\rho_{H}$ and $\rho_0$ are anomalous Hall and longitudinal resistivity correspondingly. Employing the result \eqref{eq:sublineartau}, for $T$-dependent Hall conductivity $\rho_{H}(T)$ (assuming that $\mu_v\ll T$) we arrive at
\begin{align}
    \rho_{H}(T)\propto\begin{cases}
        T^{1+1/\alpha}, & \mu_d\gg T\\
        T^{1+d+1/\alpha}, & \mu_d\ll T,
    \end{cases}
\end{align}
where $d=2,3$ is crystal dimension. One can see that deviations from $T$-linear behavior for anomalous Hall resistivity $\rho_{xy}$ could be considered as a probe of electronic band structure.

\section{Schematic Lorentz number $T$ behavior}
\begin{figure}
    \centering
    \includegraphics[width=0.45\textwidth]{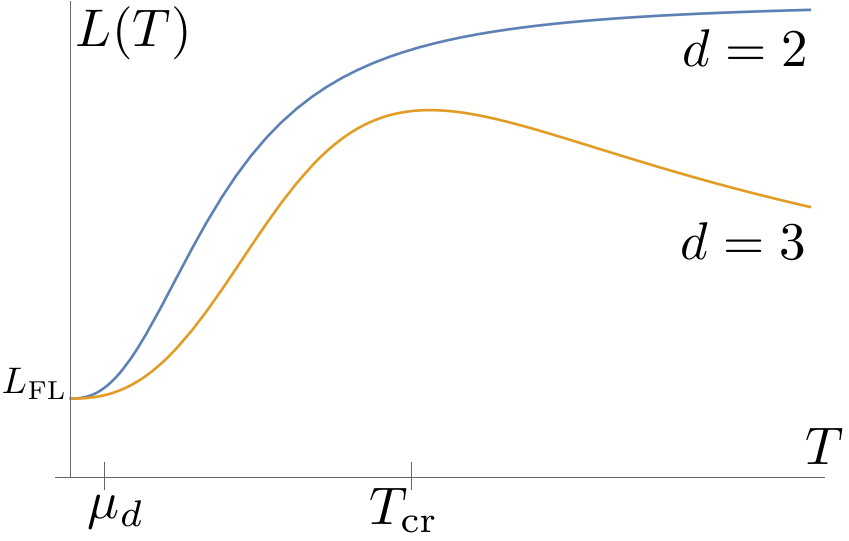}
    \caption{Lorentz number $L(T)$ versus temperature $T$. For $T\ll\mu_d$ Wiedemann-Franz law holds because both charge and heat current relax due to almost elastic internode e-e scattering. For $T\gg\mu_d$ it breaks down since although internode e-e interactions are the strongest, they do not provide thermal current relaxation for $T\gg\mu_d$. For $\mu_d<T<T_\mathrm{cr}$ thermal current relaxation is provided by internode e-e scattering which gives nontrivial power-law behavior $L(T)\propto T^\delta$, $\delta=3-3/\alpha$. For higher temperatures $T>T_\mathrm{cr}$ the increase of $L(T)$ would be cut due to energy-dependent for $T\gg\mu_d$ impurity scattering (the higher temperature behavior also depends on the dimension). $L_\mathrm{FL}=\frac{\pi^2}{3}\left(\frac{k_B}{e}\right)^2$ is Fermi liquid value of Lorentz number. $T_\mathrm{cr}$ is a crossover temperature between $\tau_\mathrm{inter}$ and $\tau_\mathrm{imp}$ dominated heat current relaxation.}
    \label{fig:Lorentz}
\end{figure}

The behavior of $L(T)$ is presented at Fig. \ref{fig:Lorentz}. Note also that even though Kagome metals are bulk materials, their band structure is still anysotropic. Therefore, for Fig. \ref{fig:Lorentz} we decided to keep both $d=2$ and $d=3$ cases, so that real material behavior should lay in between.  $T_\mathrm{cr}$ is a crossover temperature between impurity scattering-mediated and internode scattering-mediated $\tau_\kappa$, where we assumed that impurity scattering happens at short-ranged impurities, so that $\tau_\mathrm{imp}^{-1}(T)\propto \nu_d(T)$. For $d=2$ \eqref{eq:Lorentz_estimate} predicts saturation of $L(T)$ for $T>T_\mathrm{cr}$. However, this saturation behavior would switch to a decrease if one considers electron-phonon coupling at higher $T$ close to Debye temperature.

\end{document}